\documentclass[%
 aip,
 jmp,%
 amsmath,amssymb,
preprint,%
reprint,%
unsortedaddress,
]{revtex4-1}

\usepackage{graphicx}
\usepackage{dcolumn}
\usepackage{bm}

\begin{document}

\preprint{AIP/123-QED}

\title{A self-injected, diode-pumped, solid-state ring laser for laser cooling of Li atoms}

\author{Yudai Miake and Takashi Mukaiyama}
\email{muka@ils.uec.ac.jp}
\affiliation{Institute for Laser Science, University of Electro-Communications,1-5-1 Chofugaoka, Chofu, Tokyo 182-8585, Japan}

\author{Kenneth M. O'Hara}
\affiliation{Department of Physics, The Pennsylvania State University, University Park, Pennsylvania 16802-6300, USA}

\author{Stephen Gensemer}
\affiliation{Commonwealth Scientific and Industrial Research Organisation, Australia}

\date{\today}

\begin{abstract}
We have constructed a solid-state light source for experiments with laser cooled lithium atoms based on a Nd:YVO$_4$ ring laser with second-harmonic generation. Unidirectional lasing, an improved mode selection, and a high output power of the ring laser was achieved by weak coupling to an external cavity which contained the lossy elements required for single frequency operation. Continuous frequency tuning is accomplished by controlling two PZTs in the internal and the external cavities simultaneously. The light source has been utilized to trap and cool fermionic lithium atoms into the quantum degenerate regime.
\end{abstract}

\maketitle

\section{\label{sec:level1}Introduction}

Recent experimental advances in controlling inter-atomic interactions using Feshbach resonances offer great opportunities to study many-body quantum physics with the precise control of atomic physics\cite{Bloch}. Lithium atoms are particularly suitable for such study, because two isotopes of lithium with different quantum statistics (fermionic $^6$Li and bosonic $^7$Li) have relatively broad Feshbach resonances that can be utilized to precisely control inter-atomic interactions\cite{OHara, Bourdel, Zwierlein, Chin}. For Doppler cooling and an efficient trapping of the atoms, it is necessary to excite their dipole transitions with a strong laser that is red-detuned from the atomic resonance. A dye laser would be an appropriate choice of light source for cooling lithium atoms. However, dye lasers require constant care to maintain the laser condition. A combined system of a laser diode (at 671~nm) and a tapered amplifier is widely used in cold atom experiments, but these laser systems typically have a poor spatial mode quality and are not as powerful and stable compared with those for cooling other alkalis at longer resonance wavelengths. 

In this article, we report on the realization of a diode-pumped-solid-state (DPSS) light source consisting of a Nd:YVO$_4$ ring laser with second-harmonic generation (SHG) for laser cooling of lithium atoms. In previous works, single-longitudinal-mode lasers at 671~nm have been produced with the combination of a Nd:YVO$_4$ ring laser and a frequency doubler\cite{Camargo, Zheng, Eismann, Eismann2}. In these studies, unidirectional lasing was realized by placing an optical diode comprised of a terbium gallium garnet (TGG) crystal  and half-wave plate inside the ring cavity. 
The high absorption coefficient of TGG produces both a direct optical loss as well as a loss due to thermal depolarization which limits the power scaling of such ring lasers. \cite{Eismann2}.
Alternatively, one circulating mode can be favored over the other by retro-reflecting one of the two output beams back into the cavity\cite{Faxvog1, Faxvog2, Abitan, Shardlow}. However, the counter-rotating wave can never be entirely suppressed as it is required to seed the stronger of the two circulating modes. In the present work, instead of using the previously reported schemes, we realize unidirectional lasing by bringing a part of the output light back into the ring cavity through the remaining open port of the output coupler. Thus, two cavities are formed in the ring laser and an optical isolator is placed in the weakly coupled external cavity to favor one circulating direction of oscillation over the other. By placing the optical isolator in a weakly coupled external cavity, a primary contributor to the intracavity loss is eliminated. This configuration also helps to make the internal cavity length shorter to enhance the mode-hop-free tuning range.
Continuous frequency scanning can be performed by simultaneously tuning two PZTs in the internal and external cavities. The light at 671~nm is obtained by  second harmonic generation using a lithium triborate (LBO) crystal inside a doubling cavity.

\section{\label{sec:level1}laser system}

\begin{figure}[t]
\includegraphics[width=8.5cm]{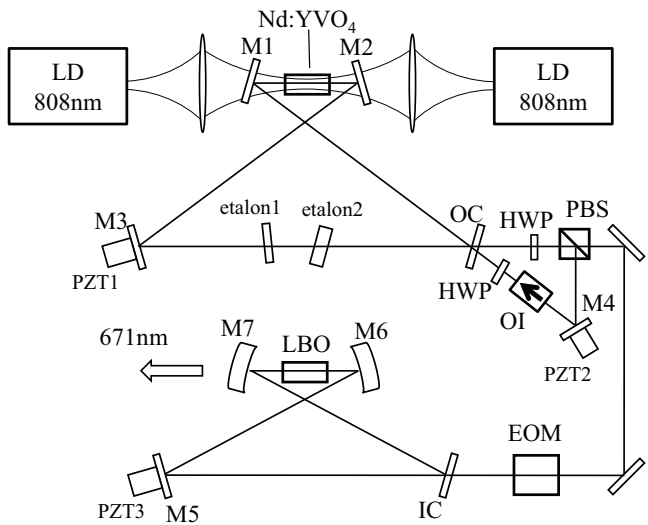}
\caption{\label{setup} Optical setup of the diode-pumped-solid-state laser at 1342~nm consisting of the fiber-coupled pump laser diodes, cavity mirrors (M1 to M4), and output coupler (OC). An optical isolator is placed together with a half-wave plate (HWP) in the external cavity path. The second cavity is used to obtain second-harmonic generation at 671~nm. LD, EOM and IC represent the laser diode, electro-optic modulator, and input coupler, respectively.}
\end{figure}

The setup of the laser system is presented in Fig. \ref{setup}. A bow-tie ring cavity is formed by four mirrors (M1, M2, M3 and OC). Two commercial fiber-coupled laser diodes (Coherent FAP-600 model) at 808~nm are used to pump the Nd:YVO$_4$ crystal through the dichroic mirrors M1 and M2 (with a high-reflection coating at 1342 nm and a high-transmission coating at 808 nm). The output fibers of the pumping laser diode have a core diameter of 600~$\mu$m, and the spot is expanded by a factor of two at the crystal position.
The Nd:YVO$_4$ crystal with a size of $3 \times 3 \times 10$~mm$^3$ is 0.15~\% doped and has anti-reflection coating at 808~nm and 1342~nm. The crystal is held by a copper mount and is temperature-controlled by a Peltier module. The OC has a reflectance of 95~\%. We intentionally use four flat mirrors (M1, M2, M3 and OC) to construct the cavity so that the stability condition of the cavity is only satisfied when the Nd:YVO$_4$ crystal causes a thermal lensing effect. Because the lowest transverse mode has the strongest thermal lensing when the pumping intensity is increased, the flat-mirror cavity helps lasing in the lowest transverse mode. Two uncoated quartz plates with different thicknesses (0.5~mm and 1~mm) are placed inside the cavity to suppress mode hops and to achieve stable single-longitudinal-mode lasing at the desired wavelength.

A part of the output (a few tens of mW) is picked by a polarization beam splitter (PBS) and is brought back into the cavity to construct another cavity after passing through an optical isolator (OI). This coupled-cavity configuration creates an asymmetry in two opposite lasing directions and achieves the unidirectional operation. Since the lasing frequency is determined both by the internal cavity (M1-M2-M3-OC-M1) and external cavity (M1-M2-M3-PBS-M4-M1) coupled to each other through the OC, the cavity lengths for both cavities need to be tuned simultaneously to sweep the laser frequency without mode hops. In this setup, two PZTs are attached to the mirrors M3 and M4 and are used to continuously sweep the laser frequency. In order to sweep the laser frequency, the voltage applied to PZT2 has to be proportional to the voltage applied to PZT1 with a factor of $(L_2-L_1)/L_1$, where $L_1$ and $L_2$ are the round-trip lengths of internal and external cavities, respectively. In our system, this factor is observed to be 0.76, which is consistent with the ratio of the cavity lengths of our actual setup. A mode-hop-free tuning range of 8~GHz at 1342~nm has been achieved, which is wider than the 3~GHz realized in the previous work \cite{Eismann2}.

\begin{figure}[t]
\includegraphics[width=8.5cm]{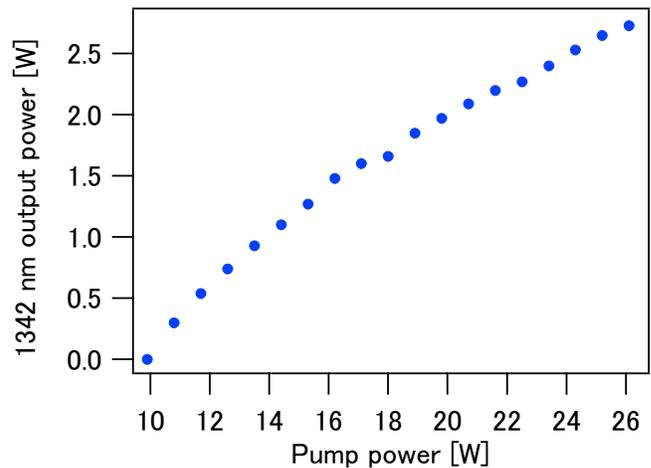}
\caption{\label{fundamental_power} DPSS output power as a function of the incident pump power. DPSS output increases with the pump power without a remarkable saturation up to 2.7~W. Slight change of the gradient is visible at a pump power of approximately 1.7~W.}
\end{figure}

\begin{figure}[b]
\includegraphics[width=8.5cm]{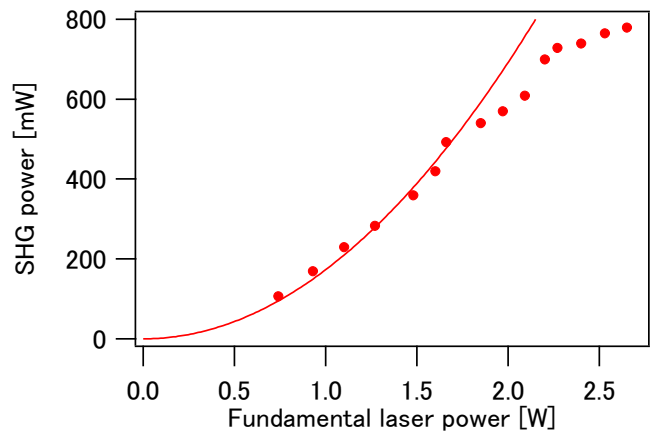}
\caption{\label{SHG_power} Closed circles show the SHG output as a function of the DPSS laser power. The solid curve shows the quadratic dependence. The plot starts to deviate from the quadratic feature approximately the 1.7~W input, indicating a slight saturation in the SHG conversion.}
\end{figure}

Figure \ref{fundamental_power} shows the output power of the 1342~nm laser as a function of the total pump power at 808~nm. The DPSS laser starts lasing at approximately 10~W of pumping and the output power increases with the pump power without a remarkable saturation; however a slight change of the gradient in the plot is observed at a pump power of 17~W. An output of 2.7~W is obtained at the maximum pumping power, which is currently limited by the cooling capacity of the Nd:YVO$_4$ crystal in our system.

To obtain laser light resonant with the atomic D2 transition of lithium atoms, the output light of the DPSS laser at 1342~nm is delivered to a cavity with an LBO crystal with a size of $3 \times 3 \times 10$~mm$^3$ inside to obtain SHG at 671~nm. The doubling cavity is frequency-locked to the free-running DPSS laser using a conventional FM sideband technique. 
The input coupler (shown as IC in Fig. \ref{setup}) has a reflectance of 97~\%; it was chosen to optimize the SHG conversion in the present experimental condition. The radius of curvature of the two concave mirrors (M6 and M7) is 50~mm. We implement the auto-relock function in this system so that the frequency lock of the doubling cavity is robust against vibration\cite{Haze}. Figure \ref{SHG_power} shows the SHG output power as a function of DPSS laser power. The SHG output increases smoothly to 800~mW with an increase in the DPSS power. The quadratic dependence of the SHG output power on the DPSS laser power (shown by the solid curve in Fig. \ref{SHG_power}) is observed up to 500~mW. The plot starts to deviate from the quadratic feature at the higher DPSS power region, indicating a saturation of the SHG conversion.

\begin{figure}[t]
\includegraphics[width=8.5cm]{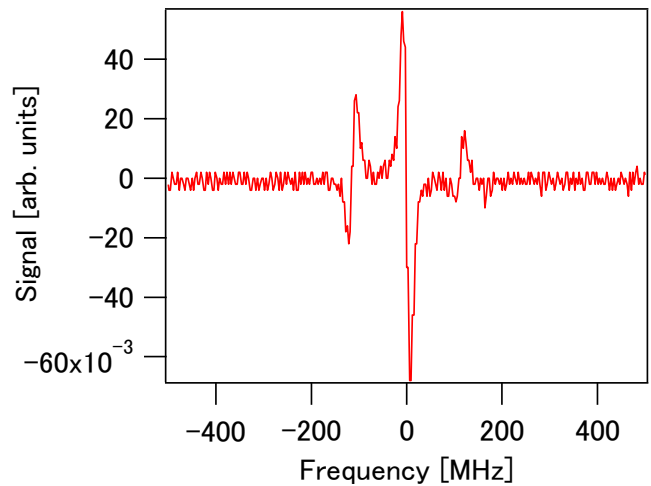}
\caption{\label{spectrum} Saturation absorption spectrum of $^6$Li atoms. Two atomic transitions (left and right resonances) and a crossover resonance signal (center) are shown. The frequency spacing between the two atomic resonances corresponds to 228~MHz, which is the hyperfine splitting of the ground state of $^6$Li. The laser is locked at the crossover resonance for cooling $^6$Li atoms in the experiment. }
\end{figure}

A few mW of SHG output is sent to a $^6$Li atomic vapor cell for saturation absorption spectroscopy. The two intra-cavity PZTs (PZT1 and PZT2) are simultaneously tuned to obtain the atomic spectrum using the frequency modulation technique (Fig. \ref{spectrum}). Three resonance signals appear in the spectrum. The left and right resonances correspond to the transitions from $^{2}S_{1/2}(F=3/2)$ and $^{2}S_{1/2}(F=1/2)$ to the excited state $^{2}P_{3/2}$, respectively. The most prominent resonance observed at the center is the crossover resonance between the two resonances. We lock the laser frequency at the crossover resonance when it is used for cold atom experiments with $^6$Li atoms. We have been able to trap $10^6$ atoms with an atomic temperature of one-tenth of the Fermi temperature, which is deeply in the Fermi-degenerate regime, using an all-optical method\cite{Inada}.

The system requires 30 minutes warmup time to reach full power and a stable lasing condition. The system requires no daily adjustment except the frequency tuning to the atomic resonance. This can be done by tuning one PZT (either PZT1 or PZT2) to set up a mode hop to bring the laser frequency within the mode-hop-free tuning range to the atomic resonance. The laser has a reasonable spatial beam quality, which provides a coupling efficency of 76 \% into a single-mode optical fiber.

We have taken the method of the feedback cavity a step further by moving the etalons from the internal cavity into the external cavity, in another demonstration.  We used this technique to stabilize a homemade dye laser for laser cooling experiments at the University of Amsterdam \cite{Tiecke}. In this instance, the external cavity contained a Faraday rotator, etalon, and a diffraction grating for further mode selection, so that the internal laser cavity contains no elements aside from the gain medium.   This can significantly reduce the length of the cavity and the cost of the optics, since much more loss can be tolerated in the external cavity.  Moreover, the addition of a diffraction grating or narrowband interference filter in the feedback path can also help to control the laser wavelength.  The technique could also be used for stabilization of a Ti:Sapphire laser, or any other ring laser cavity that normally requires intracavity mode selection.

\section{conclusion}

In summary, we have constructed a solid-state light source for an experiment in laser cooling of lithium atoms based on a Nd:YVO$_4$ ring laser followed by second-harmonic generation. 
Unidirectional lasing and improved mode selection in the ring laser was achieved by weak coupling to an external cavity containing an optical isolator. The light at 1342~nm is delivered to a doubling cavity to obtain laser light at 671~nm. Roughly 800~mW of the SHG output is obtained at the maximum pumping condition.
The output power can be further improved with higher second-harmonic-generation efficiency by preparing a higher power in the fundamental light. A primary limitation of the fundamental output power achieved to date is due to loss from the intracavity Faraday isolator and associated half waveplate. The coupled-cavity configuration presented here demonstrates a new scheme to realize unidirectional lasing while circumventing the problem of intracavity loss.
Continuous frequency tuning can be achieved by controlling two PZTs in the internal and external cavities simultaneously and the laser frequency is locked to the crossover resonance of the atomic D-line transition of $^6$Li. The light source has successfully been utilized to trap $^6$Li in the actual setup and an ultracold gas of $^6$Li deep in the quantum degenerate regime can be obtained.


\end{document}